\documentstyle[]{ptptex}

\def\Journal#1#2#3#4{{#1} {\bf #2} (#4) #3}
 
\def\PTP{Prog.Theor.Phys.}

\def\NC{Nuovo Cim.}

\def\NP{Nucl.Phys.}

\def\PL{Phys.Lett.}

\def\PRL{Phys.Rev.Lett.}

\def\PR{{Phys.Rev.}}
\def\ZP{Z.Phys.}

\def\AP{Ann. of Phys.}

\title{%
Property of ${\mib \sigma}$(600) and Chiral Symmetry
}

\author{%
Muneyuki {\sc Ishida}
}

\inst{%
Department of Physics, University of Tokyo , Tokyo 113
}

\recdate{%
March 7,1996; Revised August 20,1996
}

\abst{%
The iso-singlet scalar resonance $\sigma(600)$, 
found in the recent $\pi\pi$ phase shift analysis,
is pointed out to have the properties of the $\sigma$ meson 
in the $\sigma$ model.
The role of the $\lambda\phi^4$ interaction as 
a source of the ``background phase", 
which was essential there,
is also discussed.
}

\begin{document}
\maketitle

The chiral symmetry and the notion of its dynamical breaking have 
been playing 
a central role in understanding the spectroscopy of hadrons with 
light-quarks. Presently the non-linear representation
(Non-Linear $\sigma$ Model(NL$\sigma$M)) 
of chiral\\
$SU(2)_L\times SU(2)_R$ group
is widely accepted to be realized in nature on the basis of the 
success
of low energy theorems concerning $\pi$ mesons as 
Nambu-Goldstone bosons.

In the simplest linear representation\cite{rf:SGL}
(Linear $\sigma$ Model(L$\sigma$M)) of the chiral
symmetry, an iso-singlet scalar particle,
as $\underline{{\rm a\ chiral\ partner\ of\ }\pi}$, the $\sigma $ meson,
is required to exist in the effective Lagrangian. 
Both L$\sigma$M and NL$\sigma$M satisfy the current algebra
and PCAC, which lead to the same low-energy
$\pi\pi$ scattering amplitude $A(s,t,u)$ 
in the $O(p^2)$ level of the former in the tree-approximation
(of the latter).

The possible existence of a rather light $\sigma$ meson
has been inferred from various points of 
view.\cite{rf:Sca}${}^{\sim}$
\cite{rf:Taku} 
However, experimentally its existence has not been
accepted so far, mainly due to the negative results of 
extensive 
analyses\cite{rf:FF}${}^,$\cite{rf:MP}
of the $\pi\pi$ scattering phase shift.
The reason against its existence is that
the $I=0$ $S$-wave $\pi\pi$ phase shift $\delta_0^0$
reaches only $270^\circ$ at $m_{\pi\pi}\sim 1200$MeV,
which was considered\footnote{
However, in the newest edition of Particle Data Group'96,
the behavior of $\delta_0^0 $ is understood as due to,
in addition to the $f_0(980)$ and $f_0(1370)$, 
a very broad $f_0(400\sim 1200)$.
}
to be due mainly to the narrow $f_0(980)$
and the broad $f_0(1370)$,(leaving no room for light 
$\sigma$ particle).
In contrast with this conventional interpretation, 
the existence of a $\sigma$ resonance
has been inferred in a recent reanalysis\cite{rf:pipip} 
(referred to as I) of $\delta_0^0$.
The obtained values\footnote{
We use the values given in the revised version of I
(private communication).
}
of the resonance parameters are
\begin{eqnarray} 
m_\sigma &=& 535\sim 650{\rm MeV}, \ \  
g_{\sigma} = 1.3\sim 1.6{\rm GeV}, \ \   
\label{eq:res1}
\end{eqnarray}
where $g_{\sigma}$ is defined by
${\cal L}_{\rm int}=-g_{\sigma}\sigma{\mib \pi}^2$.
The most crucial point of obtaining this result 
is that there was an introduction
of a negative background phase of the form 
$\delta^{BG}=-p_1r_c(p_1$ being the CM pion momentum), 
analogous to
the case of the nuclear force, which implies a repulsive 
``hard-core type" interaction 
between pions.
The obtained value of $r_c$ is 
\begin{eqnarray} 
r_c &=& 0.6{\rm fm}(3.0{\rm GeV}^{-1})
   \sim 1.1{\rm fm}(5.4{\rm GeV}^{-1}).
\label{eq:res2}
\end{eqnarray}

In this paper a comment is given on the above properties of 
$\sigma$(600)
and of $\delta^{BG}$
in relation to the chiral symmetry.
The effective Lagrangian of 
L$\sigma$M\cite{rf:SGL}
is given by
\begin{eqnarray}
{\cal L}^{{\rm L}\sigma {\rm M}}  = &1/2&[(\partial_\mu{\mib \phi})^2
                              -m_\pi^2{\mib \phi}^2            
                              +(\partial_\mu\sigma ')^2
                              -m_\sigma^2\sigma '^2]
             -g_{\sigma}\sigma '({\mib \phi}^2+\sigma '^2)
             -\lambda /4({\mib \phi}^2+\sigma '^2)^2,
\nonumber\\
   &g_{\sigma}& \equiv (m_\sigma^2-m_\pi^2)/(2f_\pi ),\ \ \ \ \ \ \ \ \ \  
  \lambda\equiv (m_\sigma^2-m_\pi^2)/(2f_\pi^2).\ \ \ \ \ \ 
\label{eq:rel}
\end{eqnarray}
in terms of the physical $\sigma '$ field redefined
by $\sigma =f_\pi+\sigma '$, corresponding to the spontaneous
breakdown of chiral symmetry. 
In this ``linear form" of L$\sigma$M the
$\sigma\pi\pi$ coupling is of a non-derivative type,
and the $\pi\pi$-scattering
amplitude $A(s,t,u)$ is given by
$A^{{\rm L}\sigma {\rm M}}(s,t,u)= 
(-2g_{\sigma\pi\pi})^2/(m_\sigma^2-s)-2\lambda .
$

As was mentioned, 
the $A^{{\rm L}\sigma {\rm M}}(s,t,u)$
reduces, 
due to Eq.(\ref{eq:rel}), 
to the amplitude of NL$\sigma$M,
$A^{{\rm NL}\sigma {\rm M}}(s,t,u)= 
(s-m_\pi^2)/f_\pi^2
$
in the $O(p^2)$ approximation.
Here it is seen that, to $O(p^0)$,
the first term of $A^{{\rm L}\sigma {\rm M}}(s,t,u)$, representing the
strong attractive force proportional to $m_\sigma^2$ due to
virtual $\sigma$ production, exactly cancels the second term due to
the repulsive
$\lambda\phi^4$ interaction. This result is obtained directly
if we start from the ``non-linear form" of L$\sigma$M, where the
transformed fields $\tilde{\sigma}$ and ${\mib \pi}$ 
have an interaction of a pure derivative type.

In I,
the S matrix is represented directly through Breit-Wigner
resonance formula for $\sigma$ interacting with 
non-derivative type,\footnote{
In connection to this point, it may be notable that 
out of many phenomenological analyses, the 
ones\cite{rf:FF}${}^,$\cite{rf:MP} against the $\sigma$-existence
seem to overlook the repulsive background, 
while the \\
ones\cite{rf:BL}
${}^{\sim}$
\cite{rf:Ka2}
suggesting its existence take it into account
explicitly or implicitly, as in our case. 
}
and the amplitude ought to 
be compared with that in linear form of L$\sigma$M with
non-derivative type interaction. 

Substituting the value of $m_\sigma$ in Eq.(\ref{eq:res1}) together with 
$(m_\pi ,f_\pi )=(140,93)$MeV into Eq.(\ref{eq:rel}),
we obtain the theoretical values of $g_\sigma$ and $\lambda$ as
\begin{eqnarray}
g_{\sigma}^{\rm th.}=1.4\sim 2.2{\rm GeV},\ \ \ \ \ \ \ 
\lambda^{\rm th.}=16\sim 23,\ \ \ 
\label{eq:th}
\end{eqnarray}
of which the value $g_{\sigma}$ is 
very close to the experimental value
Eq.(\ref{eq:res1}).

The $\lambda\phi^4$ term in ${\cal L}^{{\rm L}\sigma {\rm M}}$
represents a repulsive and contact short-range interaction 
between pions
and seems to have a plausible property as an origin of $\delta^{BG}$
in the small $\sqrt{s}$-region.
The relevant I=0 amplitude consisits of a 
direct one, $A^{{\rm L}\sigma {\rm M}}(s,t,u)$, and of its crossing terms.
The crossing terms have a weak s-dependence and almost vanish
in the low energy region. Thus
the first(second) term in $A^{{\rm L}\sigma {\rm M}}(s,t,u)$ 
itself may be directly compared
to the phenomenological amplitude of 
virtual $\sigma$ production(background) in I, respectively.
The corresponding background amplitude
due to the $\lambda\phi^4$ interaction,
unitarized following the 
${\cal K}$-matrix method, gives
$\delta_{\rm th.}^{\rm BG}=-{\rm arctan}3\rho_1\lambda$,
which simulates well the behavior of the $\delta_{\rm core}^{BG}=-p_1r_c$
in the small $\sqrt{s}$-region. Then the value of $r_c$,
Eq.(\ref{eq:res2}), corresponds to the experimental value of $\lambda$,
\begin{eqnarray}
\lambda_{\rm exp}
=(p_1/3\rho_1)_{\sqrt{s}\rightarrow 2m_\pi}r_c
=16\pi m_\pi /3r_c=7\sim 13,
\label{eq:resc}
\end{eqnarray}
which is only in qualitative agreement with
the theoretical value Eq.(\ref{eq:th}b).

From the above considerations it is concluded that 
the observed properties of
$\sigma$(Eq.(\ref{eq:res1})) in I are consistent with those predicted
in the tree-approximation of L$\sigma$M,
although some discrepancy exists
concerning the $\lambda\phi^4$ term as 
a repulsive BG force.

Finally some remarks are added on the related works or problems:
The existence of $\sigma$ meson as a chiral partner of $\pi$ meson
has been strongly suggested for many years in various works
based on the NJL model or 
its extended version\cite{rf:Sca}${}^\sim$
\cite{rf:KKT}.
It is notable that the experimental value of $m_\sigma$,
Eq.(\ref{eq:res1}), obtained in I is consistent with the 
predicted one, $m_\sigma =2m_q$, in the NJL model.
In addition to this, the above result of this paper seems\footnote{
From the conventional viewpoint of NL$\sigma$M the existence of
``general $\sigma$-particle", which has no relation with 
the chiral symmetry, is also permissible.\cite{rf:EGPR}
After submitting this paper the author was informed of a work 
in which
similar analysis as in I, has been made
from this line of thought.\cite{rf:Harada}}
to suggest the validity of a linear representation
of the $\sigma$ model.
Historically, however, the L$\sigma$M, as a low energy effective
theory of hadrons, has been rejected mainly from the analyses on
the scattering lengths of all relevant Iso-spin and 
orbital momentum channels
and on the $K_{l4}$ decay form factors.\cite{rf:GL}
Corresponding to this situation I have made\footnote{
The results will be given elsewhere.
} 
reanalysis of the $\pi\pi$ scattering lengths with 
tree approximation (or in large $N_c$ limit) in L$\sigma$M,
including both effects of $\rho$- and $\sigma$-exchange,
and found rather consistent results with experiments 
(except for $a_2^2$, being somewhat larger than the experiments).
The relevant $a_0^0$ is predicted to be 
$a_0^0=0.16+0.04$, where 
the correction of higher than $O(p^4)$(the second term)
makes\footnote{Conventionally in NL$\sigma$M similar amount
of $O(p^4)$ correction is obtained which is
argued to be mainly due to
the $\rho$ meson exchange.\cite{rf:DRV}${}^,$\cite{rf:EGPR} 
Here it should be noted that
this effect does not occur in the tree level, 
and does only
due to the scale-dependence of the coefficients. 
However,  
such a large scale-dependence cannot be expected if
the $\sigma$ exists 
with fairly lower mass and larger width than $\rho$.
}
the $O(p^2)$ value(the first term) closer to the 
experimental value $a_0^0=0.26\pm 0.05$.\cite{rf:Na}
Needless to say, to be definitive, a more
extensive investigation is necessary.

\vspace*{2mm}

I would like to thank 
Professor K. Fujikawa and Professor K. Higashijima 
for instructive discussions.
I am grateful to Professor T. Yanagida 
for useful comments and encouragements.
I also thank Professor S. Ishida,T. Ishida and H. Takahashi 
for many helpful suggestions and discussions.

\end{document}